\documentclass[aps,twocolumn,pra,superscriptaddress,amsmath,showpacs,tightenlines]{revtex4-1}
\usepackage{amssymb}
\usepackage{amsmath}
\usepackage{esint}
\usepackage{graphicx}
\usepackage{subfigure}
\usepackage{natbib}
\usepackage{epsfig}
\usepackage{amsfonts}
\usepackage{mathrsfs}
\usepackage{pifont}
\usepackage{braket}  
\usepackage{xcolor}
\usepackage[toc,page,title,titletoc,header]{appendix}
\usepackage[colorlinks,linkcolor=blue,citecolor=blue]{hyperref}%
\begin{document}

\title{Liouvillian gap closing--bound states in the continuum connection and diverse dynamics in a giant-atom waveguide QED setup}

\author{Hongwei Yu}
\affiliation{Center for Quantum Sciences and School of Physics, Northeast Normal University, Changchun 130024, China}
\author{Mingzhu Weng}
\affiliation{School of Physics Science and Technology, Shenyang Normal University, Shenyang 110034, China}
\author{Zhihai Wang}
\email{wangzh761@nenu.edu.cn}
\affiliation{Center for Quantum Sciences and School of Physics, Northeast Normal University, Changchun 130024, China}
\author{Jin Wang}
\email{jin.wang.1@stonybrook.edu}
\affiliation{Department of Chemistry and of Physics and  Astronomy, Stony Brook University, Stony Brook, New York 11794-3400, USA}

\begin{abstract}
In open quantum systems, reduced dynamics is commonly described by a master equation, whose Liouvillian gap closing (LGC) typically signals the emergence of decoherence-free subspace. By contrast, the dynamics of the full system-environment compound is governed by the underlying Hamiltonian spectrum, where bound states in the continuum (BICs) can protect long-lived quantum resources. Despite these parallel perspectives, the relation between LGC and BIC formation has remained largely unexplored. Here we bridge this gap in a paradigmatic giant-atom waveguide platform and show that the occurrence of LGC necessarily benchmarks the presence of a BIC in the full Hamiltonian description. By engineering the giant-atom geometry, we further demonstrate rich dynamical regimes-including Rabi oscillations, fractional decay, and complete exponential relaxation-depending on the number of supported BICs, which can be tuned from three to zero. Remarkably, when two BICs become frequency-degenerate, the long-time dynamics approaches a steady state rather than exhibiting persistent oscillations. Our results establish a direct spectral-dynamical connection between effective Markovian and underlying non-Markovian descriptions, and provide a route toward flexible control of open-system dynamics.
 \end{abstract}

\maketitle
\section{introduction}

Emerging quantum technologies are intimately connected to the dynamics of open quantum systems, with prominent examples including precision sensing in quantum metrology~\cite{SA2014,VM2025}, quantum information processing~\cite{FV2009,AV2022}, and quantum thermal machines~\cite{JP2021,JP2015}. Consequently, achieving controllable and robust manipulation of open-system dynamics has become a central goal in modern quantum physics.

In the theory of open quantum systems, one is often interested in the reduced dynamics of a target system obtained by tracing out environmental degrees of freedom. A widely used description relies on the Markov approximation, where the environment is assumed to be memoryless, leading to a time-local master equation of the form $d\rho/dt=\mathcal{L}\rho$ with $\rho$ the reduced density matrix and $\mathcal{L}$ a generally non-Hermitian Liouvillian superoperator~\cite{HW2010,SH2006,HJ1999}. The spectral properties of $\mathcal{L}$ govern the relaxation and decoherence of the reduced dynamics~\cite{FM2018,SL2025,JZ2025,YL2023,FM2019}. In particular, the appearance of multiple eigenmodes with vanishing real parts-often referred to as Liouvillian gap closing (LGC)-signals the emergence of decoherence-free (or noiseless) subspaces and has been widely exploited in quantum information processing~\cite{HR2024,BB2012,BB2019,VV2014,ZZ2020,CB2020}.

Beyond the Markovian regime, however, the dynamics is ultimately determined by the spectrum of the full system-environment Hamiltonian. The environment typically provides a continuum in the energy (frequency) domain, within which a bound state in the continuum (BIC)~\cite{FH1975,DC2008,MI2012,GC2019} may arise: a state localized in real space while embedded in the continuous spectrum, enabled by destructive-interference mechanisms~\cite{HF1985,SF2003,SW2013,MV2017,SI2018,LL2024,ER2024}. Such BICs, which can be also predicted by the equations of movement, have been shown to protect long-lived quantum resources, including coherence and entanglement~\cite{AA2011,AK2017,BZ2013}, in non-Markovian settings. Despite these parallel insights, the connection between the LGC in effective Markovian descriptions and the existence of BICs in the underlying Hamiltonian picture has received comparatively little attention~\cite{an2016}, leaving an apparent gap between Markovian and non-Markovian perspectives.

As a representative setting, we consider three giant atoms~\cite{MV2014,AF2021,WZ2020,AF2014,ST2022,XWT2021,XWZ2022,AS2022,NL2022,XWH2022,CJ2023,LG2017,CA2021,LG2020,SG2020,XL2022,LD2023,XZ2023,GA2019} coupled to a coupled resonator waveguide (CRW) and use this model to address the above issue. In this example, the CRW acts as the environment and provides an energy continuum through its cosine-type dispersion relation, whereas the giant atoms-whose spatial extent is comparable to the photon wavelength and thus beyond the dipole approximation~\cite{DF2008}-constitute the open quantum system of interest. We show that the occurrence of LGC always implies the presence of BIC in the underlying system-environment Hamiltonian.

Moreover, by engineering the geometry of the giant atoms-namely, their sizes and relative positions-we can tune the number of BICs supported by the composite system from three to zero, thereby accessing a variety of atomic dynamical behaviors. When three BICs are present, we observe coherent Rabi oscillations~\cite{SL2021,KH2023,CW2016,MF2017,KK2019,KKG2019}; when only one BIC exists, the dynamics exhibits fractional decay~\cite{JT2022,ST2006,PL2000}; and when no BIC is supported, the excitation undergoes a complete exponential decay, as also found in other physical platforms~\cite{WA2008,HG1996,FL2014}.

Interestingly, when the system supports two BICs, we find that the long-time dynamics approaches a steady state rather than displaying persistent oscillations. We attribute this nontrivial steady behavior to the degeneracy of the two BICs together with the selected initial conditions. This regime is in sharp contrast to the long-time oscillations associated with two BICs reported in recent studies~\cite{SL2021,HY2025,AS2023,KH2023,ER2024}. Finally, we provide an analytical treatment that captures the BIC-induced nonexponential relaxation and reproduces the dynamical regimes described above.


The remainder of this paper is organized as follows. In Sec.~\ref{model}, we introduce the model of three giant atoms coupled to a CRW. In Sec.~\ref{Liouvillian gap closing}, we establish the connection between LGC and the existence of BICs. In Sec.~\ref{Dynamics behaviors}, we present the resulting dynamical behaviors of the giant-atom system and show how they depend on the number of supported BICs. Finally, Sec.~\ref{Conclusion} summarizes our main findings. Technical details, including the derivations of the master equation and the probability-amplitude evolution equations, are provided in the Appendices.

\section{Model}
\label{model}
\begin{figure}
\begin{centering}
\includegraphics[width=1.0\columnwidth]{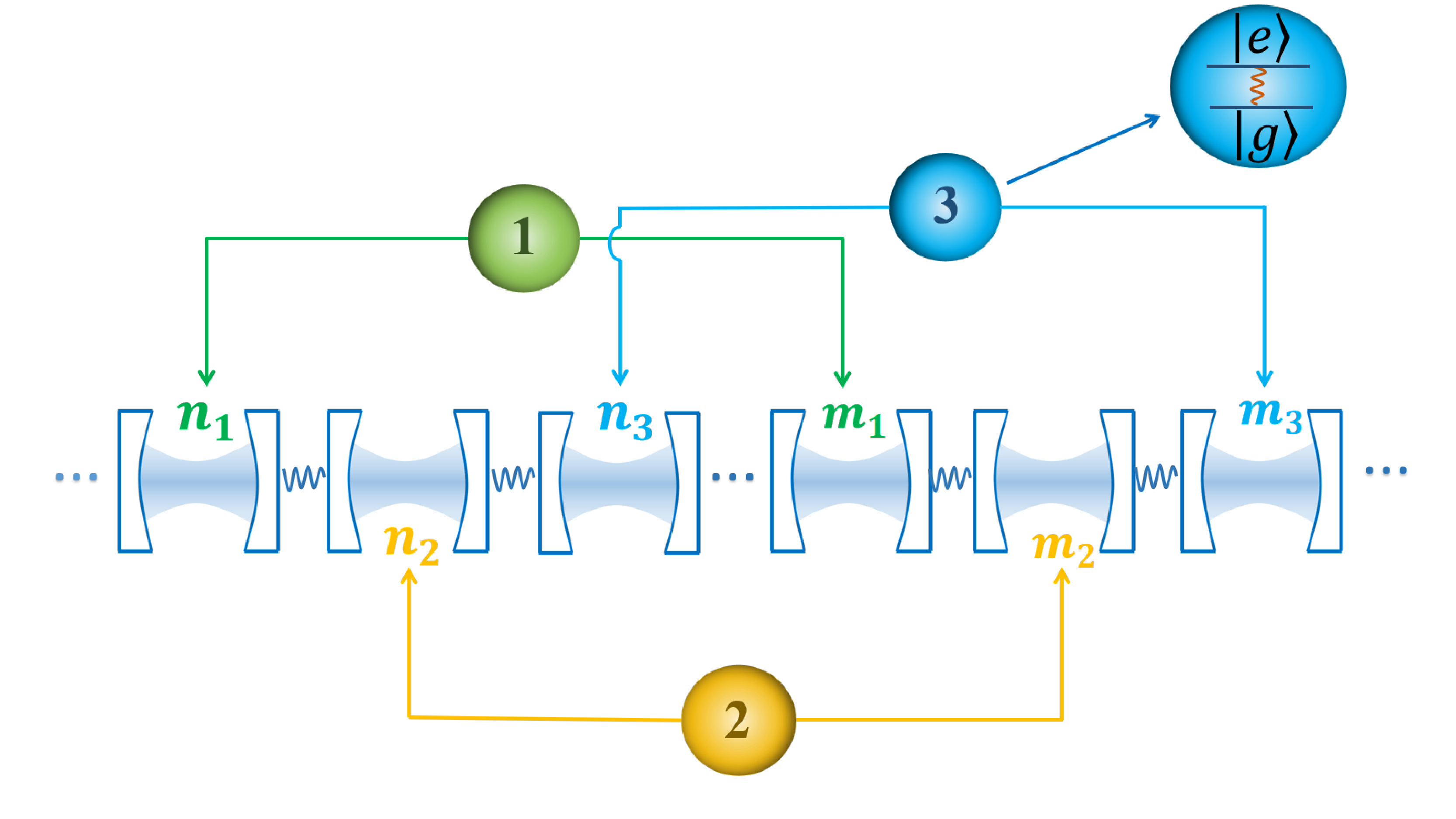}
\par\end{centering}
\caption{Schematic of three braided giant atoms coupled to a CRW. The $i$th giant atom couples to the waveguide at two sites, labeled $n_i$ and $m_i$. Throughout this work, we consider equal-size giant atoms and impose a braided geometry for any pair of atoms, which is ensured by the condition $n_1<n_2<n_3<m_1$.}
\label{setup}
\end{figure}

As shown in Fig.~\ref{setup}, we consider a system of three braided giant atoms coupled to a one-dimensional coupled-resonator waveguide (CRW) at two spatially separated sites for each atom. The total Hamiltonian can be written as
$H=H_{a}+H_{c}+H_{I}$ (we set $\hbar=1$ throughout),
\begin{eqnarray}
H_{a} & = & \sum_{i=1}^{3}\Omega_{i}\left|e\right\rangle _{i}\left\langle e\right|,\\
H_{c} & = & \omega_{c}\sum_{j}a_{j}^{\dagger}a_{j}-\xi\sum_{j}\left(a_{j+1}^{\dagger}a_{j}+a_{j}^{\dagger}a_{j+1}\right),\\
H_{I} & = & \sum_{i=1}^{3}g_{i}\left(a_{n_{i}}^{\dagger}\sigma_{i}^{-}+a_{m_{i}}^{\dagger}\sigma_{i}^{-}+\mathrm{H\text{.c.}}\right).
\end{eqnarray}
Here, $H_{a}$ denotes the free Hamiltonian of the atomic subsystem, where $\Omega_{i}$ is the transition frequency of the $i$th giant atom between the excited state $\left|e\right\rangle$ and the ground state $\left|g\right\rangle$. The CRW Hamiltonian $H_{c}$ describes an array of single-mode resonators with bare frequency $\omega_{c}$ and nearest-neighbor hopping strength $\xi$, where $a_{j}^{\dagger}$ ($a_{j}$) creates (annihilates) a photon in the $j$th resonator. The interaction Hamiltonian $H_{I}$ accounts for the coupling between the giant atoms and the CRW: $g_{i}$ is the coupling strength of the $i$th atom, $n_i$ and $m_i$ label the positions of its two coupling points, and $\sigma_{i}^{-}=|g\rangle_i\langle e|$ is the corresponding lowering operator.

In the limit of an infinite CRW, $N_{c}\rightarrow\infty$, it is convenient to introduce the Fourier transform
$a_{j}=\frac{1}{\sqrt{N_{c}}}\sum_{k}e^{ikj}a_{k}$~\cite{SL2020},
which diagonalizes the waveguide Hamiltonian as $H_{c}=\sum_{k}\omega_{k}a_{k}^{\dagger}a_{k}$, with the dispersion relation
$\omega_{k}=\omega_{c}-2\xi\cos k,\,k\in[-\pi,\pi)$.
The corresponding energy band is centered at $\omega_{c}$ and has a bandwidth $4\xi$, providing a structured photonic environment.
Accordingly, the total Hamiltonian in momentum space reads
\begin{eqnarray}
H & = & \sum_{i=1}^{3}\Omega_{i}\left|e\right\rangle _{i}\left\langle e\right|
+\sum_{k}\omega_{k}a_{k}^{\dagger}a_{k}\nonumber \\
 &  & +\sum_{i=1}^{3}\sum_{k}\frac{g_{i}}{\sqrt{N_{c}}}
\left[\left(e^{ikn_{i}}+e^{ikm_{i}}\right)a_{k}^{\dagger}\sigma_{i}^{-}
+\mathrm{H\text{.c.}}\right].
\end{eqnarray}

\section{Liouvillian gap closing and bound state in the continuum}
\label{Liouvillian gap closing}

\subsection{Markov Approximation: Liouvillian Gap}
\label{Markovian dynamics equations}
Treating the waveguide as an environment, we obtain, within the Markov approximation, a master equation for the reduced density matrix of the atomic subsystem (see Appendix~\ref{A} for details),
\begin{eqnarray}
\label{ME}
\dot{\rho_{a}}
 & = & -i\left[\sum_{i=1}^{3}\Omega_{i}\left|e\right\rangle _{i}\left\langle e\right|,\rho_a\right]\nonumber \\
 &  & +\sum_{i\text{,}j=1}^{3}A_{ij}\left(\sigma_{j}^{-}\rho_a\sigma_{i}^{+}-\sigma_{i}^{+}\sigma_{j}^{-}\rho_a\right)\nonumber \\
 &  & +A_{ij}^{*}\left(\sigma_{i}^{-}\rho_a\sigma_{j}^{+}-\rho_a\sigma_{j}^{+}\sigma_{i}^{-}\right).
\end{eqnarray}
The coefficients $A_{ij}$ take the form
\begin{eqnarray}
A_{ij} & = & \frac{g_{i}g_{j}}{2\xi}\left(e^{i\frac{\pi}{2}\left|n_{i}-n_{j}\right|}
+e^{i\frac{\pi}{2}\left|n_{i}-m_{j}\right|}\right.\nonumber \\
 &  & \left.+e^{i\frac{\pi}{2}\left|m_{i}-n_{j}\right|}
+e^{i\frac{\pi}{2}\left|m_{i}-m_{j}\right|}\right).
\end{eqnarray}
Since Eq.~(\ref{ME}) is linear in $\rho_a$, it can be cast in the compact form~\cite{FM2018} $\dot{\rho_{a}}=\mathcal{L}\rho_a$,
where the time-independent superoperator $\mathcal{L}$ is the Liouvillian. The real parts of its eigenvalues, $\mathrm{Re}[\lambda_{i}]$, determine the relaxation rates toward the steady state.

\begin{figure}[tbp]
\centering
\includegraphics[width=\columnwidth]{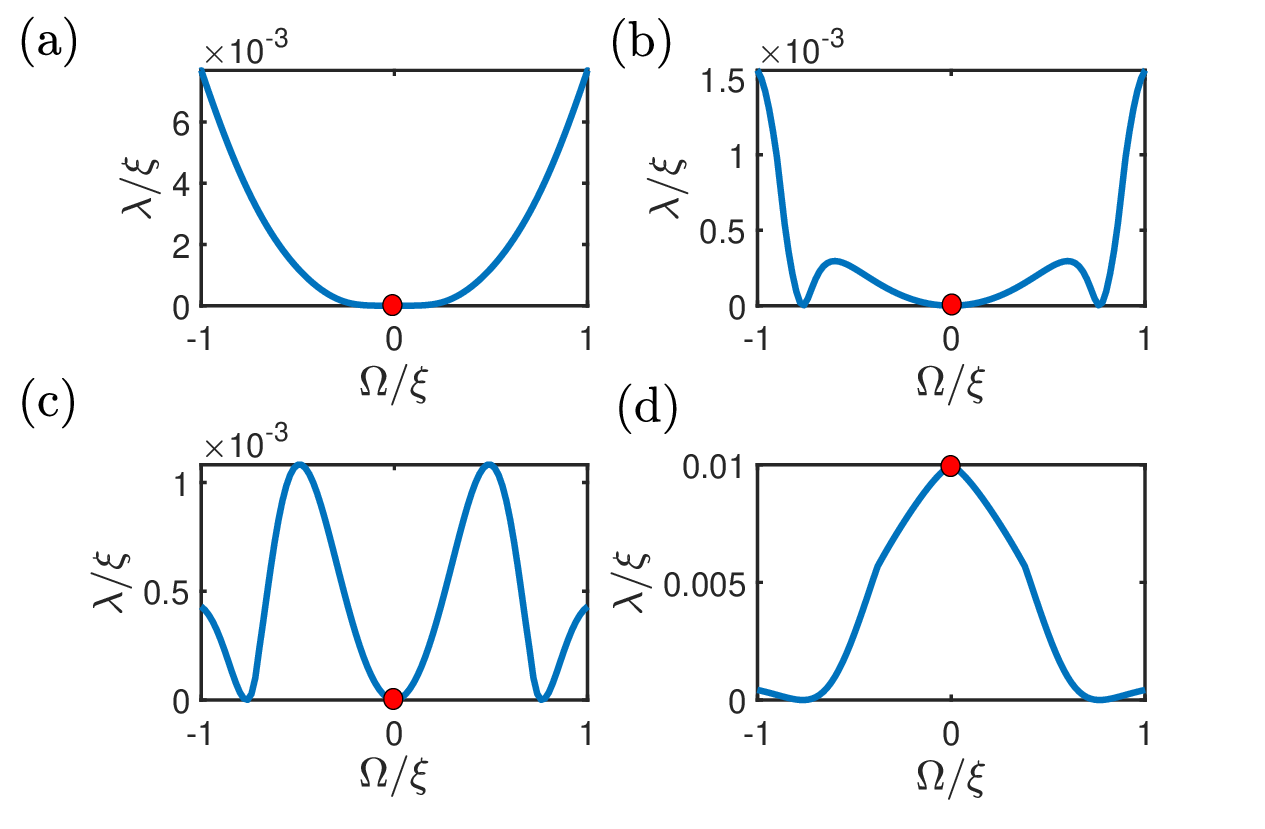}
\vspace{8pt}
\setlength{\tabcolsep}{7.5pt}
\hspace{14pt}
\begin{tabular}{c c c c c c c}
\hline\hline
      & $n_1$ & $m_1$ & $n_2$ & $m_2$ & $n_3$ & $m_3$ \\
\hline
(a)   & 1     & 7     & 2     & 8     & 4     & 10    \\
(b)   & 1     & 9     & 5     & 13    & 7     & 15    \\
(c)   & 1     & 9     & 5     & 13    & 6     & 14    \\
(d)   & 1     & 9     & 2     & 10    & 3     & 11    \\
\hline\hline
\end{tabular}
\caption{Liouvillian gap $\lambda$ as a function of the atomic frequency $\Omega$. The parameters are
$\Omega_{1}=\Omega_{2}=\Omega_{3}=\Omega$, $\omega_{c}=0$, and $g_{1}=g_{2}=g_{3}=0.1\xi$.
The table specifies the atomic configuration in each panel.}
\label{gap}
\end{figure}

It has been shown that $\mathrm{Re}\left[\lambda_{i}\right]\leq 0$ for all $i$~\cite{HP2007,AR2011}. For convenience, we order the eigenvalues such that
$\left|\mathrm{Re}\left[\lambda_{0}\right]\right|<\left|\mathrm{Re}\left[\lambda_{1}\right]\right|<\cdots<\left|\mathrm{Re}\left[\lambda_{n}\right]\right|$,
where $\lambda_{0}=0$ always exists and the corresponding eigenstate describes a steady state~\cite{DE1997,BB2008,AR2012}.
The Liouvillian gap is then defined as $\lambda=\left|\mathrm{Re}\left[\lambda_{1}\right]\right|$, which sets the slowest relaxation rate in the long-time limit~\cite{EM2012}. In particular, $\lambda=0$, that is, LGC, indicates a complete suppression of complete relaxation (i.e., dissipation is quenched).

For our setup, by tuning the giant-atom geometry---namely, their sizes and relative positions---we find that the dependence of the Liouvillian gap on the atomic frequency can be grouped into four representative classes, as shown in Fig.~\ref{gap}.
Throughout this analysis, we assume three resonant giant atoms and choose the bare cavity frequency as the reference point, i.e., $\Omega_1=\Omega_2=\Omega_3=\Omega$ and $\omega_c=0$.
Under these conditions,  LGC is observed whenever $\Omega=0$ in Figs.~\ref{gap}(a)--(c), corresponding to the geometries listed in the first three rows of the table in the bottom panel.
In contrast, for the geometry given in the last row, no LGC is found, as shown in Fig.~\ref{gap}(d). For any other geometries, we find that they always obey one of the above four results.

\begin{figure}
\centering
\includegraphics[width=1\columnwidth]{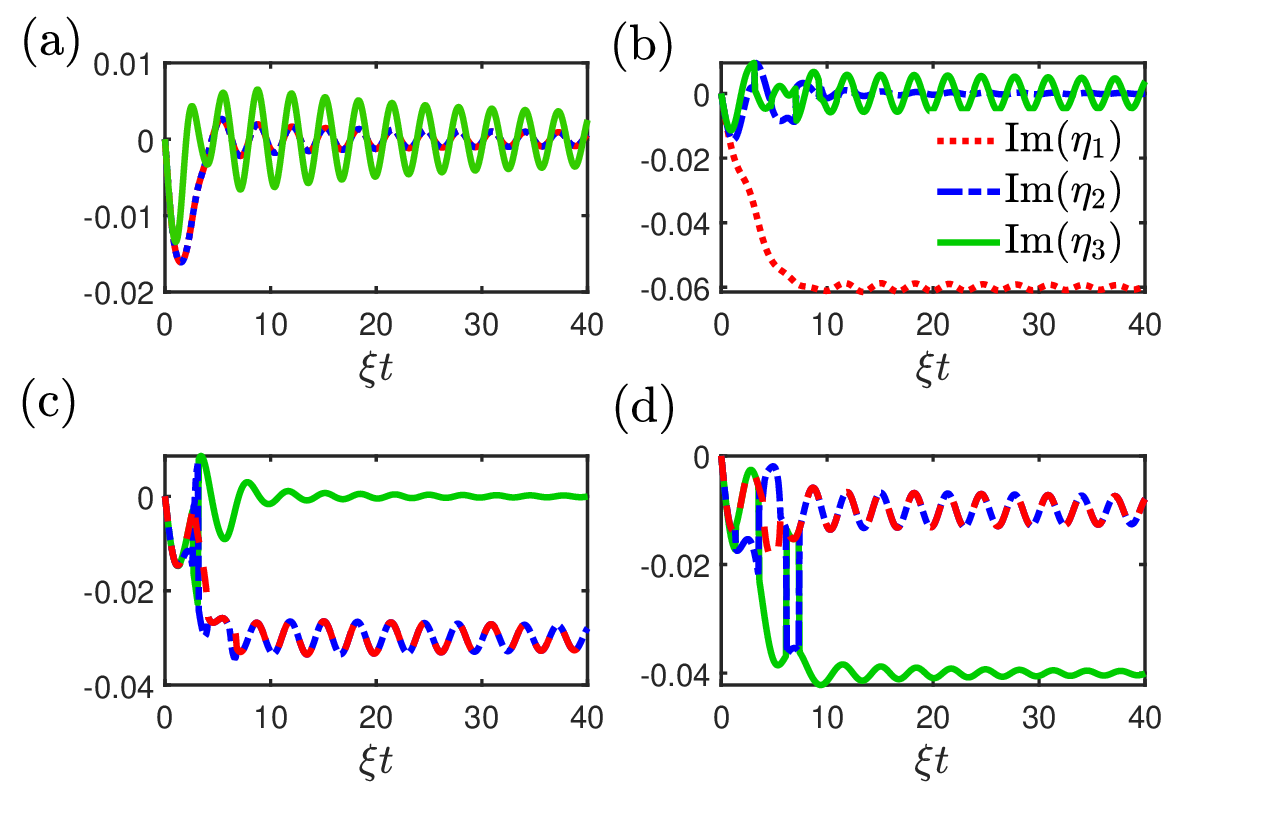}
\caption{Imaginary parts of the three eigenvalues of the matrix $M(t)$. (a) Three BICs. (b) Two BICs. (c) One BIC. (d) No BIC. Parameters are
$\Omega_{1}=\Omega_{2}=\Omega_{3}=\omega_{c}=0$ and $g_{1}=g_{2}=g_{3}=0.1\xi$.
The atomic configuration for each panel is given in the table in Fig.~\ref{gap}.}
\label{selfenergy}
\end{figure}

\subsection{Beyond the Markov Approximation: BICs}
\label{Beyond Markovian dynamics equations}
Starting from the total Hamiltonian, we derive the dynamics of  the atomic and photonic degrees of freedom. In what follows, we focus on the single-excitation manifold, for which the time-dependent state can be written as
\begin{equation}
\label{wave1}
\left|\psi\left(t\right)\right\rangle
=\left[\sum_{i=1}^{3}\alpha_{i}\left(t\right)\sigma_{i}^{+}
+\sum_{k}\beta_{k}\left(t\right)a_{k}^{\dagger}\right]\left|G\right\rangle.
\end{equation}
Here, $\left|G\right\rangle$ denotes the state in which all atoms are in the ground state and the waveguide is in the vacuum. The coefficient $\alpha_{i}\left(t\right)$ is the probability amplitude for the $i$th atom to be excited, while $\beta_{k}\left(t\right)$ is the probability amplitude for occupying the $k$th waveguide mode.
Within the Weisskopf--Wigner approach, while going beyond the Markov approximation, the dynamics of the three giant atoms can be cast as (see Appendix~\ref{B} for details)
\begin{equation}
\label{atom}
i\frac{\partial}{\partial t}\left(\begin{array}{c}
\alpha_{1}\left(t\right)\\
\alpha_{2}\left(t\right)\\
\alpha_{3}\left(t\right)
\end{array}\right)
=M\left(t\right)\left(\begin{array}{c}
\alpha_{1}\left(t\right)\\
\alpha_{2}\left(t\right)\\
\alpha_{3}\left(t\right)
\end{array}\right),
\end{equation}
where the time-dependent matrix $M\left(t\right)$ has the form
\begin{equation}
M\left(t\right)=\left(\begin{array}{ccc}
\mathcal{A}_{1}\left(t\right) & \mathcal{B}_{12}\left(t\right) & \mathcal{B}_{13}\left(t\right)\\
\mathcal{B}_{12}\left(t\right) & \mathcal{A}_{2}\left(t\right) & \mathcal{B}_{23}\left(t\right)\\
\mathcal{B}_{13}\left(t\right) & \mathcal{B}_{23}\left(t\right) & \mathcal{A}_{3}\left(t\right)
\end{array}\right).
\end{equation}
The diagonal elements describe the self-energy contributions, which take the form
\begin{eqnarray}
\mathcal{A}_{i}\left(t\right) & = & \Omega_{i}-2ig_{i}^{2}\int_{0}^{t}d\tau\, e^{-i\omega_{c}\tau}\nonumber \\
 &  & \times\left\{ J_{0}\left(2\xi\tau\right)+i^{\left|N_{i}\right|}J_{\left|N_{i}\right|}\left(2\xi\tau\right)\right\}.
\end{eqnarray}
Here, $N_i=m_i-n_i$ characterizes the size of the $i$th giant atom, and $J_{|m|}$ denotes the Bessel function of the first kind of order $|m|$.
The off-diagonal element $\mathcal{B}_{ij}\left(t\right)$ represents the waveguide-mediated interaction between the $i$th and $j$th atoms, given by
\begin{eqnarray}
\mathcal{B}_{ij}\left(t\right) & = & -ig_{i}g_{j}\int_{0}^{t}d\tau\, e^{-i\omega_{c}\tau}\nonumber \\
 &  & \times \sum_{p,q=n,m}\left\{ i^{\left|p_{i}-q_{j}\right|}J_{\left|p_{i}-q_{j}\right|}\left(2\xi\tau\right)\right\}.
\end{eqnarray}

For the two-giant-atom case, the number of BICs equals the number of eigenvalues $\eta$ of the matrix $M$ whose imaginary parts vanish, $\mathrm{Im}(\eta)=0$~\cite{SL2021}. The same criterion applies to the three-giant-atom configuration considered here. The underlying reason is that $\mathrm{Im}(\eta)$ quantifies the radiative loss of the corresponding collective atomic mode into the waveguide. Owing to the coupling to the environment, these imaginary parts are nonpositive, i.e., $\mathrm{Im}(\eta)\le 0$.
When $\mathrm{Im}(\eta)=0$, the associated mode does not decay into the CRW and is effectively decoupled from the waveguide, thereby forming a BIC. Figure~\ref{selfenergy} shows the time evolution of $\mathrm{Im}(\eta)$. In Fig.~\ref{selfenergy}(a), all three eigenvalues satisfy $\mathrm{Im}(\eta_{1})=\mathrm{Im}(\eta_{2})\approx 0$ and $\mathrm{Im}(\eta_{3})\approx 0$, indicating three BICs in the full system. In Fig.~\ref{selfenergy}(b), $\mathrm{Im}(\eta_{2})\approx 0$ and $\mathrm{Im}(\eta_{3})\approx 0$ while $\mathrm{Im}(\eta_{1})<0$, corresponding to two BICs. In Fig.~\ref{selfenergy}(c), only $\mathrm{Im}(\eta_{3})\approx 0$ whereas $\mathrm{Im}(\eta_{1})=\mathrm{Im}(\eta_{2})<0$, yielding a single BIC. Finally, in Fig.~\ref{selfenergy}(d), all three eigenvalues have nonzero imaginary parts, $\mathrm{Im}(\eta_{1}),\mathrm{Im}(\eta_{2}),\mathrm{Im}(\eta_{3})\neq 0$, implying that no BIC is supported.

In Fig.~\ref{selfenergy}, we set the atomic frequency to $\Omega=\omega_{c}=0$ (red dots), and the other parameters in each panel are identical to those used in Fig.~\ref{gap}, respectively. By comparing Figs.~\ref{gap}(a)--(c) and \ref{selfenergy}(a)--(c), we observe that $\lambda=0$ coincides with the emergence of BICs in the full system. In contrast, comparing Fig.~\ref{gap} (d) with Fig.~\ref{selfenergy}(d) shows that when $\lambda\neq 0$ no BIC is supported. Therefore, in our setup  the Liouvillian gap closing ($\lambda=0$) provides a direct signature of the presence of BICs.

\begin{figure*}
\centering
\includegraphics[width=0.9\columnwidth]{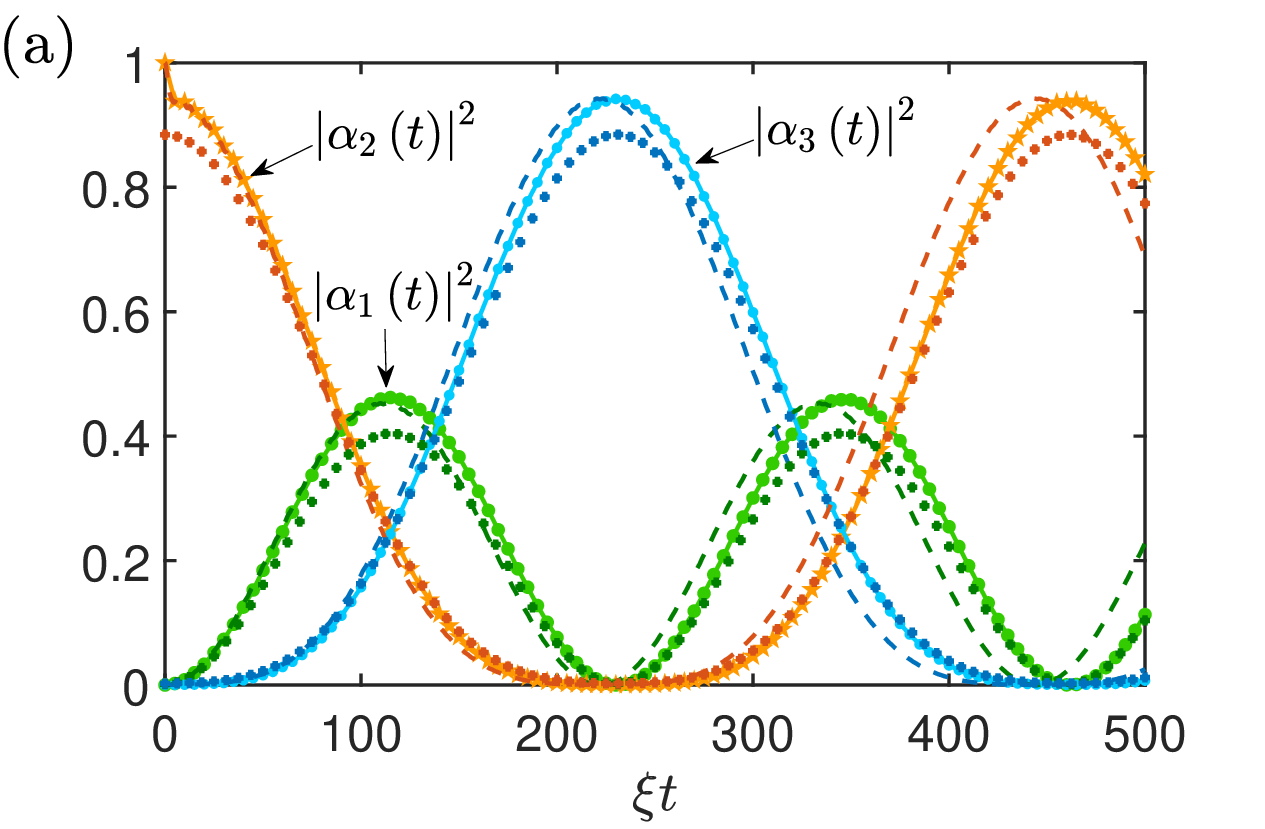}
\includegraphics[width=0.9\columnwidth]{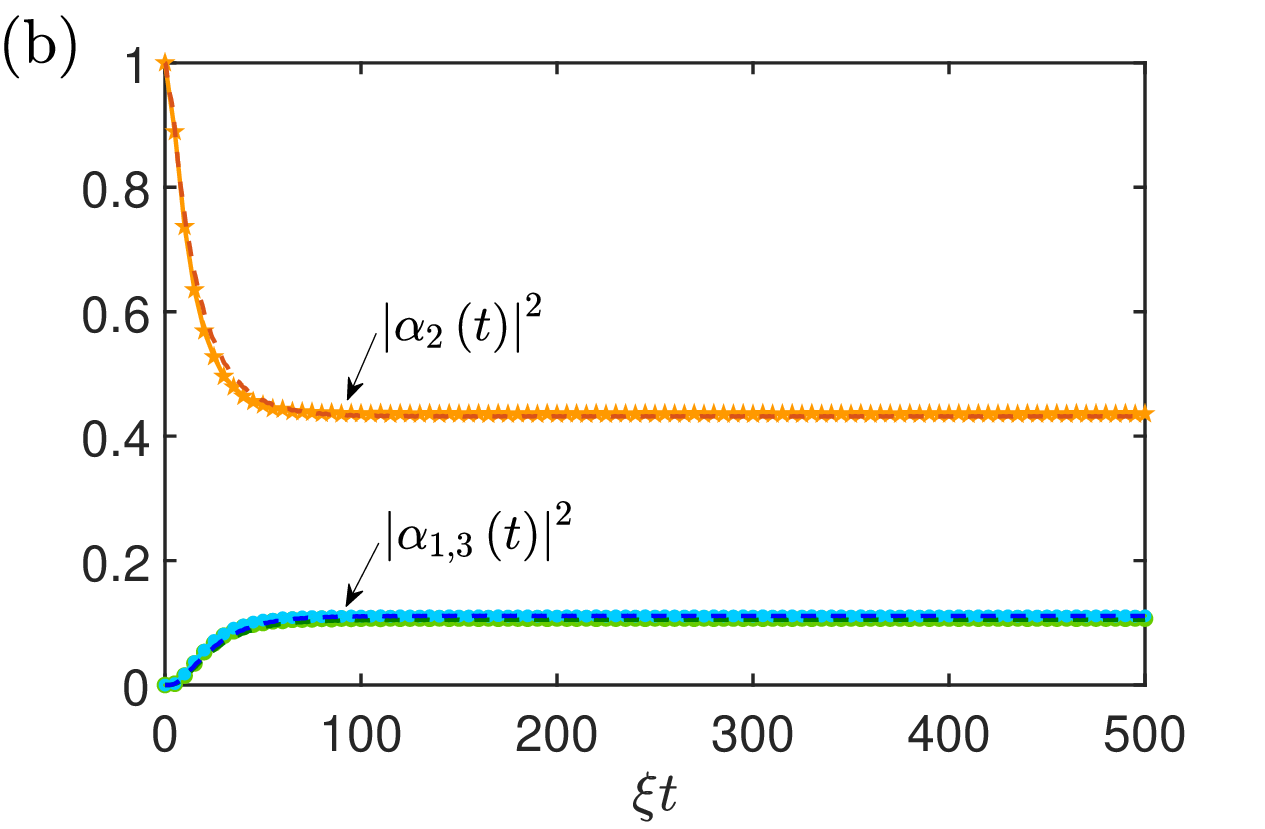}
\includegraphics[width=0.9\columnwidth]{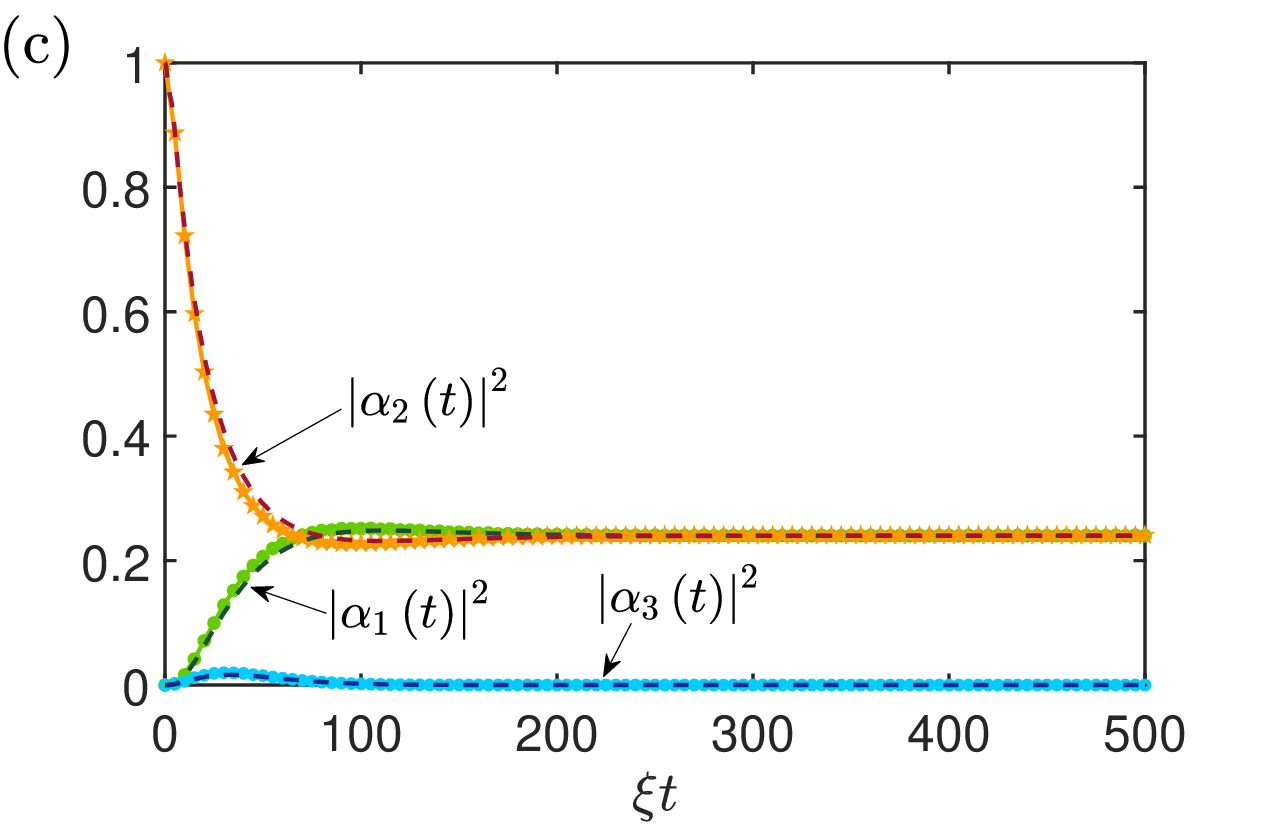}
\includegraphics[width=0.9\columnwidth]{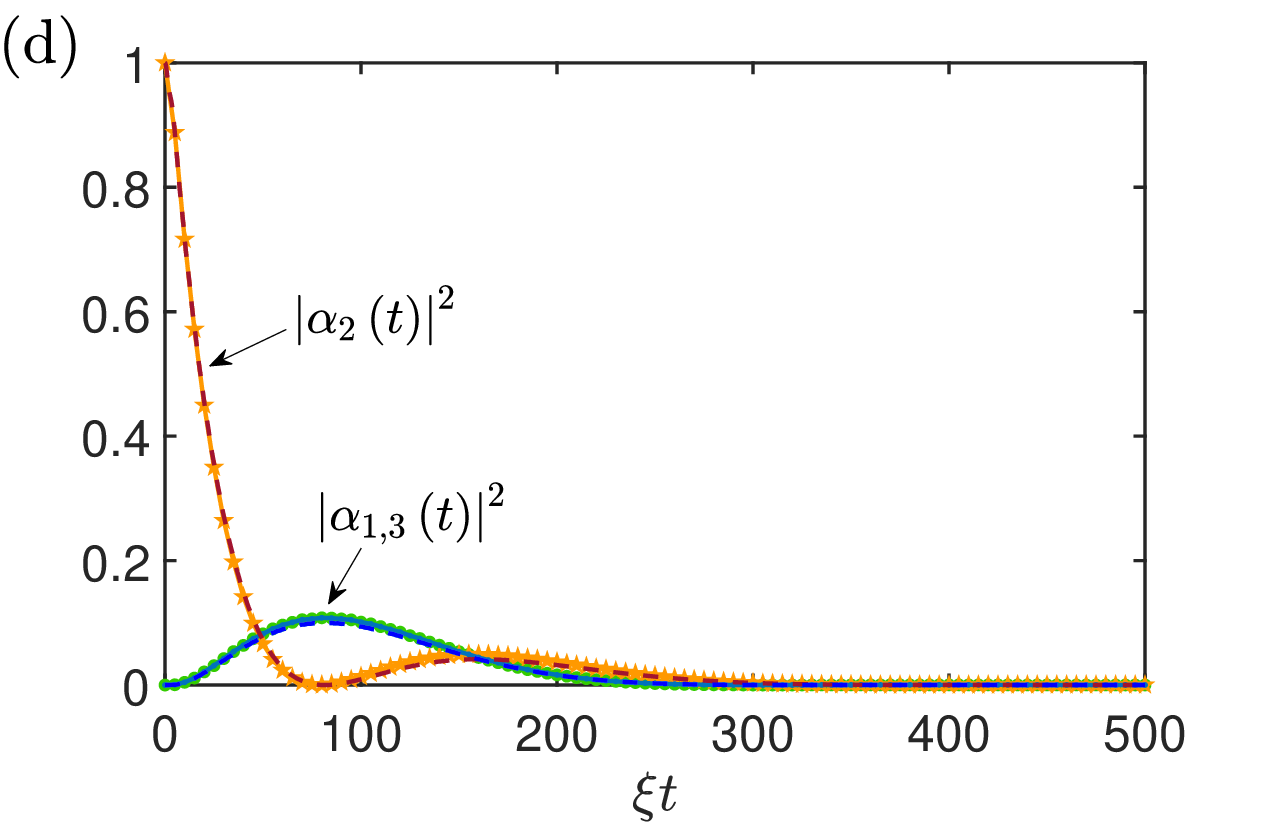}
\caption{Population dynamics of the three giant atoms. (a) Three BICs. (b) Two BICs. (c) One BIC. (d) No BIC. Parameters are
$\Omega_{1}=\Omega_{2}=\Omega_{3}=\omega_{c}=0$ and $g_{1}=g_{2}=g_{3}=0.1\xi$.
The atomic configuration for each panel is given in the table in Fig.~\ref{gap}.}
\label{Pe}
\end{figure*}

\section{Diverse Dynamical Behaviors}
\label{Dynamics behaviors}
In this section, we explore the distinct dynamical behaviors associated with different numbers of BICs. For a single giant atom of size $N=6$ coupled to the CRW, the global system supports one BIC, and an initially excited atom does not decay~\cite{XJ2023}. When three such atoms are combined, three BICs emerge, independent of the relative positions among the atoms. By contrast, for a single giant atom of size $N=8$, no BIC exists and an initially excited atom eventually relaxes to the ground state~\cite{XJ2023}. Interestingly, when three $N=8$ atoms are combined, the number of BICs becomes sensitive to their relative positions, and the system can host two, one, or zero BICs, as demonstrated in the last section. In the following, we present the dynamical results for these four representative scenarios in turn.

\subsection{Three BICs}

To illustrate the case of three BICs, we choose, as a representative example,
$n_1=1,m_1=7$, $n_2=2,m_2=8$, and $n_3=4,m_3=10$.
Starting from the initial state $\left|\psi\left(0\right)\right\rangle=\sigma_{2}^{+}\left|G\right\rangle$,
we plot in Fig.~\ref{Pe}(a) the population dynamics of the three atoms,
$\left|\alpha_{1}\left(t\right)\right|^{2}$, $\left|\alpha_{2}\left(t\right)\right|^{2}$, and $\left|\alpha_{3}\left(t\right)\right|^{2}$.
The solid curves show the numerically exact results obtained from
$\left|\psi\left(t\right)\right\rangle=e^{-iHt}\left|\psi\left(0\right)\right\rangle$,
whereas the dashed curves correspond to the analytical solution of Eq.~(\ref{atom}).
The agreement between the two confirms the validity of the Weisskopf--Wigner treatment in the present parameter regime.
We find that the populations exhibit pronounced oscillations, indicating a persistent exchange of excitation between the different atomic subsystem.
This behavior also demonstrates that, despite the absence of any direct atom--atom coupling, the CRW mediates an effective interaction among the three giant atoms.

Under this configuration, the full system supports three BICs. Denoting these BICs by
$\ket{\varphi_{B}^{(n)}}$ with eigenenergies $E_{B}^{(n)}$ ($n=1,2,3$), we numerically confirm that the long-time oscillations originate from the coherent superposition of these three BICs. Specifically, in the long-time limit the atomic populations are well captured by
\begin{eqnarray}
\left|\alpha_{i}\left(t\rightarrow\infty\right)\right|^{2}
& = &
\left|\sum_{n=1}^{3}e^{-iE_{B}^{\left(n\right)}t}
\left\langle G\left|\sigma_{i}^{+}\right|\varphi_{B}^{\left(n\right)}\right\rangle
\right.\nonumber\\
& &
\left.
\times
\left\langle \varphi_{B}^{\left(n\right)}\left|\psi\left(0\right)\right.\right\rangle
\right|^{2},
\label{superposition}
\end{eqnarray}
which yields the dotted curves in Fig.~\ref{Pe}(a).
The overlap $\langle G|\sigma_{i}^{+}|\varphi_{B}^{(n)}\rangle$ determines which pairs of BICs contribute to the oscillations of the $i$th atomic population; the corresponding values are summarized in Table~\ref{projection}, where the $i$th column lists the projection of each BIC onto the excited state of the $i$th atom.
From the first column, we see that the oscillation in $|\alpha_{1}(t)|^{2}$ is dominated by the interference between $\ket{\varphi_{B}^{(1)}}$ and $|\varphi_{B}^{(3)}\rangle$, with period
$T_{1}=2\pi/|E_{B}^{(1)}-E_{B}^{(3)}|$.
From the second column, each $\ket{\varphi_{B}^{(n)}}$ has a nonzero projection onto the excitation of atom~2; consistent with Fig.~\ref{Pe}(a), the resulting dynamics involves two beating components associated with the pairs
$|\varphi_{B}^{(1)}\rangle \leftrightarrow |\varphi_{B}^{(2)}\rangle$ and
$|\varphi_{B}^{(3)}\rangle \leftrightarrow |\varphi_{B}^{(2)}\rangle$,
with characteristic periods
$T_{2}=2\pi/|E_{B}^{(1,3)}-E_{B}^{(2)}|$.
Similarly, the third column shows that the behavior of atom~3 is symmetric to that of atom~2, reflecting the corresponding Rabi-type exchange between these two atoms, as also seen in Fig.~\ref{Pe}(a).

\begin{table}[tbp]
\centering
\caption{Overlaps $\langle G|\sigma_{i}^{+}|\varphi_{B}^{(n)}\rangle$. Parameters are the same as in Fig.~\ref{Pe}(a).}
\label{projection}
\renewcommand{\arraystretch}{1.2}
\begin{tabular*}{\columnwidth}{@{\extracolsep{\fill}} c c c c}
\hline\hline
$n$ &
$\langle G|\sigma_{1}^{+}|\varphi_{B}^{(n)}\rangle$ &
$\langle G|\sigma_{2}^{+}|\varphi_{B}^{(n)}\rangle$ &
$\langle G|\sigma_{3}^{+}|\varphi_{B}^{(n)}\rangle$ \\
\hline
1 & $0.654$ & $-0.459$ & $0.459$ \\
2 & $0$     & $0.704$  & $0.704$ \\
3 & $0.693$ & $0.485$  & $-0.485$ \\
\hline\hline
\end{tabular*}
\end{table}

\subsection{Two BICs}
To illustrate the case of two BICs, we choose $n_1=1,m_1=9$, $n_2=5,m_2=13$, and $n_3=7,m_3=15$.
We again consider the initial state $\left|\psi\left(0\right)\right\rangle=\sigma_{2}^{+}\left|G\right\rangle$.
As shown in Fig.~\ref{Pe}(b), the excitation of atom~2 does not decay to zero but instead approaches a finite population.
Meanwhile, atoms~1 and~3 are also populated and retain nonzero steady-state excitations, indicating that dissipation in the atomic subsystem is strongly suppressed.
In this regime, the long-time state is well described by a coherent superposition of the two BICs, $\ket{\varphi_{B}^{(n)}}$ ($n=1,2$), with eigenenergies $E_{B}^{(n)}$, yielding
\begin{eqnarray}
\left|\alpha_{i}\left(t\rightarrow\infty\right)\right|^{2}
& = &
\left|\sum_{n=1}^{2}e^{-iE_{B}^{\left(n\right)}t}
\left\langle G\left|\sigma_{i}^{+}\right|\varphi_{B}^{\left(n\right)}\right\rangle
\right.\nonumber\\
& &
\left.
\times
\left\langle \varphi_{B}^{\left(n\right)}\left|\psi\left(0\right)\right.\right\rangle
\right|^{2}.
\end{eqnarray}
We numerically confirm that this expression reproduces the steady-state values in Fig.~\ref{Pe}(b).
Notably, because the two BICs are energy-degenerate, $E_{B}^{(1)}=E_{B}^{(2)}$, the atomic dynamics do not exhibit persistent oscillations, and the steady states population is obtained as
\begin{eqnarray}
\left|\alpha_{i}\left(t\rightarrow\infty\right)\right|^{2} & = & \sum_{n=1}^{2}\left|\left\langle G\left|\sigma_{i}^{+}\right.\left|\varphi_{B}^{\left(n\right)}\right.\right\rangle \left\langle \left.\varphi_{B}^{\left(n\right)}\right|\psi\left(0\right)\right\rangle \right|^{2}\nonumber \\
 & + & 2\mathrm{Re}\left[\prod_{n=1}^{2}\left\langle G\left|\sigma_{i}^{+}\right.\left|\varphi_{B}^{\left(n\right)}\right.\right\rangle \left\langle \left.\varphi_{B}^{\left(n\right)}\right|\psi\left(0\right)\right\rangle \right].\nonumber \\
\end{eqnarray}
This behavior goes beyond the common expectation that two BICs necessarily generate long-lived Rabi oscillations~\cite{SL2021,HY2025,AS2023,KH2023,ER2024}.

\subsection{One BIC}
To illustrate the case of a single BIC, we choose $n_1=1,m_1=9$, $n_2=5,m_2=13$, and $n_3=6,m_3=14$.
As shown in Fig.~\ref{Pe}(c), we again take the initial state $\left|\psi\left(0\right)\right\rangle=\sigma_{2}^{+}\left|G\right\rangle$.
The excitation does not decay completely to zero: atom~1 and 2 are populated to a finite value, whereas atom~3 remains in the ground state.
In the long-time limit, the system relaxes into the BIC $\ket{\varphi_{B}}$, and the residual populations are given by
\begin{equation}
\left|\alpha_{i}\left(t\rightarrow\infty\right)\right|^{2}
=\left|\left\langle G\left|\sigma_{i}^{+}\right|\varphi_{B}\right\rangle
\left\langle \varphi_{B}\left|\psi\left(0\right)\right.\right\rangle\right|^{2}.
\end{equation}
This indicates that the initial state contains a finite overlap with the BIC as well as with extended states. As time evolves, the contributions from the extended states are washed out by destructive interference, leaving only the nondecaying BIC component.

\subsection{No BIC}
Owing to the cosine dispersion of the CRW, the corresponding spectral density is approximately flat when the atomic transition frequency lies near the band center. In this weak-coupling regime, an atom coupled to the waveguide undergoes essentially Markovian spontaneous emission with a standard exponential decay~\cite{FL2014}.
In our system, the same behavior arises whenever no BIC is supported. As a representative example, we choose
$n_1=1,m_1=9$, $n_2=2,m_2=10$, and $n_3=3,m_3=11$.
Starting again from $\left|\psi\left(0\right)\right\rangle=\sigma_{2}^{+}\left|G\right\rangle$, the excited-state population decays exponentially, as shown in Fig.~\ref{Pe}(d), and all atoms relax to the ground state at long times.
This behavior reflects the absence of photonic localization: without BICs, the emitted photons are not trapped but instead propagate away and spread throughout the waveguide. Physically, the CRW then acts as an effectively memoryless environment that induces irreversible dissipation of the atomic subsystem, without appreciable information backflow.

\section{Conclusion}
\label{Conclusion}

In conclusion, we have established an initial connection between LGC and the emergence of BICs in a paradigmatic giant-atom waveguide QED setup, where three giant atoms are coupled to a coupled-resonator waveguide. This platform is naturally accessible in superconducting-circuit architectures: transmon qubits can serve as giant atoms~\cite{AF2018,BK2020,GA2019,AM2021,ZW2022} with characteristic frequencies in the few-gigahertz range, and their coupling to superconducting resonators~\cite{MJ2006,DG2007,AA2012,AD2010} can reach the hundreds-of-megahertz regime, comparable to the photon-hopping rate of the waveguide.

Our results show that LGC within an effective Markovian treatment may benchmarks the presence of BICs in the full system--environment Hamiltonian, thereby enabling rich non-Markovian dynamics. In particular, when the spectrum supports three BICs, two frequency-degenerate BICs or a single BIC, the system can exhibit persistent energy exchange and steady-state trapping, respectively, offering routes to protect and manipulate quantum information.

More broadly, our work can be straightforwardly extended to more complex open-system settings, paving the way toward a unified framework in which Markovian and non-Markovian descriptions are mutually compatible.

\section*{acknowledgments}
This work is supported by National Natural Science Foundation of China (Grant No. 12375010) and Quantum Science and Technology-National Science and Technology Major Project (No. 2023ZD0300700).

\section*{Data availability}

The data that support the findings of
this paper are not publicly and are available from the authors upon reasonable request.

\appendix
\addcontentsline{toc}{section}{Appendices}\markboth{APPENDICES}{}

\begin{subappendices}
\begin{widetext}

\section{Markovian Master Equation}
\label{A}

In this appendix, we derive the master equation for the atomic subsystem within the Markov approximation. In the interaction picture, the reduced density operator of an open system obeys the formal equation
\begin{equation}
\dot{\rho}(t)=-\int_{0}^{\infty}d\tau\,\mathrm{Tr}_{c}\!\left\{
\left[H_{I}(t),\left[H_{I}(t-\tau),\rho_{c}\otimes\rho(t)\right]\right]\right\}.
\end{equation}
In the interaction picture, the atom--waveguide coupling reads
\begin{equation}
H_{I}(t)=\sum_{i=1}^{3}g_{i}\left[\sigma_{i}^{-}E^{\dagger}\!\left(n_{i},m_{i},t\right)e^{-i\Omega_{i}t}
+\mathrm{H.c.}\right],
\end{equation}
where
$E^{\dagger}\!\left(n_{i},m_{i},t\right)=\frac{1}{\sqrt{N_{c}}}\sum_{k}\left(e^{ikn_{i}}+e^{ikm_{i}}\right)a_{k}^{\dagger}e^{i\omega_{k}t}$,
and $n_i,m_i$ label the two coupling points of the $i$th giant atom.
At zero temperature, the waveguide is initially in the vacuum state, such that
$\mathrm{Tr}_{\rm c}\!\left[E^{\dagger}\!\left(n_{i},m_{i},t\right)E\!\left(n_{j},m_{j},t-\tau\right)\rho_{c}\right]=0$.
Tracing out the waveguide degrees of freedom and transforming back to the Schr\"{o}dinger picture, we obtain the master equation for the three giant atoms,
\begin{equation}
\dot{\rho}_{a}=-i\left[H_{a},\rho_{a}\right]
+\sum_{i,j=1}^{3}A_{ij}\left(\sigma_{j}^{-}\rho_{a}\sigma_{i}^{+}-\sigma_{i}^{+}\sigma_{j}^{-}\rho_{a}\right)
+A_{ij}^{*}\left(\sigma_{i}^{-}\rho_{a}\sigma_{j}^{+}-\rho_{a}\sigma_{j}^{+}\sigma_{i}^{-}\right),
\end{equation}
with~\cite{GC2016}
\begin{eqnarray}
A_{ij} & = & g_{i}g_{j}\int_{0}^{\infty}d\tau\, e^{i\Omega_{i}\tau}\,
\mathrm{Tr}_{c}\!\left[E\!\left(n_{i},m_{i},t\right)E^{\dagger}\!\left(n_{j},m_{j},t-\tau\right)\rho_{c}\right]\nonumber \\
 & = & g_{i}g_{j}\int_{0}^{\infty}d\tau\,\frac{e^{i\Omega_{i}\tau}}{N_{c}}\,
\mathrm{Tr}_{c}\!\left[\sum_{kk'}e^{-i\omega_{k}t}\left(e^{-ikn_{i}}+e^{-ikm_{i}}\right)a_{k}\,
e^{i\omega_{k'}\left(t-\tau\right)}\left(e^{ikn_{j}}+e^{ikm_{j}}\right)a_{k'}^{\dagger}\rho_{c}\right]\nonumber \\
 & = & g_{i}g_{j}\int_{0}^{\infty}d\tau\,\frac{1}{N_{c}}\sum_{k}e^{-i\left(\omega_{k}-\Omega_{i}\right)\tau}
\left(e^{-ikn_{i}}+e^{-ikm_{i}}\right)\left(e^{ikn_{j}}+e^{ikm_{j}}\right).
\end{eqnarray}

Expanding the expression for $A_{ij}$ yields four contributions. As an illustration, we evaluate one of them,
\begin{eqnarray}
A_{ij}^{\left(1\right)}
& = & g_{i}g_{j}\int_{0}^{\infty}d\tau\,\frac{1}{N_{c}}\sum_{k}
e^{-i\left(\omega_{k}-\Omega_{i}\right)\tau}e^{-ik\left(n_{j}-n_{i}\right)}\nonumber \\
& = & g_{i}g_{j}\int_{0}^{\infty}d\tau\,\frac{1}{N_{c}}\sum_{n=0}^{N_{c}-1}
e^{-i\left(\omega_{c}-\Omega_{i}\right)\tau}\,
e^{-2\pi i\left(n_{j}-n_{i}\right)n/N_{c}}\,
e^{2i\xi\cos\left(2\pi n/N_{c}\right)\tau}\nonumber \\
& = & g_{i}g_{j}\int_{0}^{\infty}d\tau\,\frac{e^{-i\left(\omega_{c}-\Omega_{i}\right)\tau}}{N_{c}}
\sum_{n=0}^{N_{c}-1}e^{-2\pi i\left(n_{j}-n_{i}\right)n/N_{c}}
\sum_{m=-\infty}^{\infty}i^{m}J_{m}\left(2\xi\tau\right)e^{2\pi inm/N_{c}}\nonumber \\
& = & g_{i}g_{j}\int_{0}^{\infty}d\tau\, e^{-i\left(\omega_{c}-\Omega_{i}\right)\tau}\,
i^{\left|n_{i}-n_{j}\right|}J_{\left|n_{i}-n_{j}\right|}\left(2\xi\tau\right)\nonumber \\
& = & \frac{g_{i}g_{j}e^{iK\left|n_{i}-n_{j}\right|}}{\sqrt{4\xi^{2}-\left(\omega_{c}-\Omega_{i}\right)^{2}}},
\end{eqnarray}
where $K=\pi-\arccos\!\left[(\Omega_{i}-\omega_{c})/(2\xi)\right]$.
In the second line we used the discretization $k=2\pi n/N_c$, and in the third line we applied the Jacobi--Anger expansion
$e^{iz\cos\theta}=\sum_{m=-\infty}^{\infty} i^{m}J_{m}(z)e^{im\theta}$.
In the last step, we employed the integral identity leading to the closed form above. In particular, for the resonant case
$\Omega_{1}=\Omega_{2}=\Omega_{3}=\omega_{c}$, one has $K=\pi/2$.
For completeness, we also use the standard integral
\begin{equation}
\int_{0}^{\infty}d\tau\, J_{m}\left(a\tau\right)=\frac{1}{\left|a\right|},
\end{equation}
which holds under the usual conditions on $a$.
Evaluating the remaining three contributions in the same way, we obtain
\begin{equation}
A_{ij}=\frac{g_{i}g_{j}}{2\xi}\left(
e^{i\frac{\pi}{2}\left|n_{i}-n_{j}\right|}
+e^{i\frac{\pi}{2}\left|n_{i}-m_{j}\right|}
+e^{i\frac{\pi}{2}\left|m_{i}-n_{j}\right|}
+e^{i\frac{\pi}{2}\left|m_{i}-m_{j}\right|}
\right).
\end{equation}

\section{Dynamics Beyond the Markov Approximation}
\label{B}
In this appendix, we provide a detailed derivation of the dynamical equations beyond the Markov approximation. In momentum space, the total Hamiltonian reads
\begin{eqnarray}
\label{toalhamition}
H=\sum_{i=1}^{3}\Omega_{i}\left|e\right\rangle _{i}\left\langle e\right|
+\sum_{k}\omega_{k}a_{k}^{\dagger}a_{k}
+\sum_{i=1}^{3}\sum_{k}\frac{g_{i}}{\sqrt{N_{c}}}\left[\left(e^{ikn_{i}}+e^{ikm_{i}}\right)a_{k}^{\dagger}\sigma_{i}^{-}
+\mathrm{H\text{.c.}}\right].
\end{eqnarray}
Restricting to the single-excitation manifold, we write the time-dependent state as
\begin{equation}
\left|\psi\left(t\right)\right\rangle
=\left[\sum_{i=1}^{3}\alpha_{i}\left(t\right)\sigma_{i}^{+}
+\sum_{k}\beta_{k}\left(t\right)a_{k}^{\dagger}\right]\left|G\right\rangle.
\end{equation}
Substituting this ansatz into the Schr\"{o}dinger equation, $i\partial_t\ket{\psi(t)}=H\ket{\psi(t)}$, yields
\begin{eqnarray}
\label{alpha}
i\frac{\partial}{\partial t}\alpha_{i}\left(t\right)
& = & \Omega_{i}\alpha_{i}\left(t\right)
+\sum_{k}\frac{g_{i}}{\sqrt{N_{c}}}\left(e^{-ikn_{i}}+e^{-ikm_{i}}\right)\beta_{k}\left(t\right),\\
i\frac{\partial}{\partial t}\beta_{k}\left(t\right)
& = & \omega_{k}\beta_{k}\left(t\right)
+\sum_{i=1}^{3}\frac{g_{i}}{\sqrt{N_{c}}}\left(e^{ikn_{i}}+e^{ikm_{i}}\right)\alpha_{i}\left(t\right).
\label{beta}
\end{eqnarray}

Here we assume that the waveguide is initially in the vacuum state, i.e., $\beta_{k}(0)=0$. Equation~(\ref{beta}) then gives
\begin{equation}
\label{betak}
\beta_{k}(t)=-i\sum_{i=1}^{3}\frac{g_{i}}{\sqrt{N_{c}}}\left(e^{ikn_{i}}+e^{ikm_{i}}\right)
\int_{0}^{t}d\tau\,\alpha_{i}(\tau)e^{-i\omega_{k}(t-\tau)}.
\end{equation}
Substituting Eq.~(\ref{betak}) into Eq.~(\ref{alpha}) and using the identities
\begin{equation}
e^{iz\cos\theta}=\sum_{n=-\infty}^{+\infty}i^{n}J_{n}(z)e^{in\theta},\qquad
\int_{-\pi}^{\pi}e^{i(n-m)k}\,dk=2\pi\delta_{n,m},\qquad
\frac{1}{\pi}\int_{-\pi}^{\pi}dk\,e^{2i\xi\cos k\,(t-\tau)}=2J_{0}\!\left[2\xi(t-\tau)\right],
\end{equation}
together with $J_{-N}(x)=(-1)^{N}J_{N}(x)$, we obtain
\begin{eqnarray}
\frac{\partial}{\partial t}\alpha_{i}(t)
& = & -i\Omega_{i}\alpha_{i}(t)
-\sum_{j=1}^{3}g_{i}g_{j}\int_{0}^{t}d\tau\,\alpha_{j}(\tau)\,
e^{-i\omega_{c}(t-\tau)}
\sum_{p,q=n,m}\left\{ i^{\left|p_{i}-q_{j}\right|}
J_{\left|p_{i}-q_{j}\right|}\!\left[2\xi(t-\tau)\right]\right\}.
\end{eqnarray}
Next, within the Weisskopf--Wigner approximation we replace $\alpha_{j}(\tau)$ by $\alpha_{j}(t)$, yielding
\begin{eqnarray}
\frac{\partial}{\partial t}\alpha_{i}(t)
& = & -i\Omega_{i}\alpha_{i}(t)
-\sum_{j=1}^{3}g_{i}g_{j}\alpha_{j}(t)\int_{0}^{t}d\tau\, e^{-i\omega_{c}\tau}
\sum_{p,q=n,m}\left\{ i^{\left|p_{i}-q_{j}\right|}
J_{\left|p_{i}-q_{j}\right|}\!\left(2\xi\tau\right)\right\}.
\end{eqnarray}
For clarity, the above set of equations can be cast into the matrix form of Eq.~(\ref{atom}) in the main text.
\end{widetext}
\end{subappendices}


\end{document}